\documentclass[preprint,prd,aps,tighten,nofootinbib,amssymb]{revtex4}
\usepackage{graphicx}
\usepackage{epsfig,amssymb,amsmath}
\usepackage{slashed, color}

\newcommand{\beq}{\begin{equation}}
\newcommand{\eeq}{\end{equation}}
\newcommand{\bea}{\begin{eqnarray}}
\newcommand{\eea}{\end{eqnarray}}
\newcommand{\bear}{\begin{array}}
\newcommand {\eear}{\end{array}}
\newcommand{\bef}{\begin{figure}}
\newcommand {\eef}{\end{figure}}
\newcommand{\bec}{\begin{center}}
\newcommand {\eec}{\end{center}}

\newcommand{\dis}[1]{\begin{equation}\begin{split}#1\end{split}\end{equation}}

\newcommand{\lsim}{
\mathrel{\hbox{\rlap{\hbox{\lower4pt\hbox{$\sim$}}}\hbox{$<$}}}}
\newcommand{\gsim}{
\mathrel{\hbox{\rlap{\hbox{\lower4pt\hbox{$\sim$}}}\hbox{$>$}}}}

\newcommand{\Mpl}{M_{\rm Pl}}
\newcommand{\Lbr}{\Lambda_{\rm b}}
\newcommand{\Lbrm}{\Lambda_{\rm max}}

\begin{document}
\draft
\tighten
\preprint{CTPU-16-29}

\title{\large \bf Constraints on Relaxion Windows}

\author{
    Kiwoon Choi$^{a}$\footnote{email: kchoi@ibs.re.kr},
    Sang Hui Im$^{b}$\footnote{email: shim@th.physik.uni-bonn.de },
}
     
\affiliation{
$^a$Center for Theoretical Physics of the Universe, \\
 Institute for Basic Science (IBS), Daejeon 34051, Korea \\
 $^b$Bethe Center for Theoretical Physics and Physikalisches Institut der Universit\"at Bonn \\
 Nussallee 12, 53115 Bonn, Germany \\
    }

\vspace{4cm}

\begin{abstract}
We examine the low energy phenomenology of the relaxion solution to the weak scale hierarchy problem.
Assuming that the Hubble friction is responsible for a dissipation of the relaxion energy,   we identify the cosmological relaxion window which corresponds to the parameter region compatible with a given value of the acceptable number of inflationary $e$-foldings.
We then discuss a variety of observational constraints on the relaxion window, including those from astrophysical and cosmological considerations.
We find that majority of the parameter space with a relaxion mass  $m_\phi\gtrsim 100$ eV or a relaxion decay constant  $f\lesssim 10^7$ GeV is excluded by existing constraints. There is an interesting parameter region with $m_\phi\sim \,0.2-10$ GeV and $f\sim\, {\rm few}-200$ TeV, which is allowed by existing constraints, but can be probed soon by future beam dump experiments such as the SHiP experiment, or by improved EDM experiments.
\end{abstract}

\maketitle

\section{Introduction
\label{sec:introduction}}


Recently cosmological relaxation of the Higgs boson mass has been proposed as a new solution to the weak scale hierarchy problem
\cite{Graham:2015cka},  leading to a number of subsequent works to explore its viability \cite{Espinosa:2015eda, Antipin:2015jia, Gupta:2015uea, Patil:2015oxa, Jaeckel:2015txa, Batell:2015fma, Matsedonskyi:2015xta, Marzola:2015dia, DiChiara:2015euo, Ibanez:2015fcv, Fonseca:2016eoo, Fowlie:2016jlx, Evans:2016htp, Huang:2016dhp}\footnote{For a similar earlier idea, see Ref. \cite{Dvali:2003br}.}.
The scheme involves an axion-like field, the relaxion $\phi$, 
which scans the Higgs boson mass in the early universe from an initial value 
comparable to the cutoff scale $M\gg v=246$ GeV  to the final value of ${\cal O}(v)$. Such cosmological relaxation of the Higgs boson mass can be achieved by assuming the following form of the relaxion potential: 
\dis{
V(\phi, h) = \mu_h^2 (\phi) |h|^2 + V_0(\phi) + V_{\rm b}(\phi, h),
}
where $h$ is the Standard Model (SM) Higgs field and
\bea
\mu_h^2(\phi) &=& M^2 -  M^2 \frac{\phi}{f_{\rm eff}} + \cdots , \\
V_0(\phi) &=& -c_0 M^4 \frac{\phi}{f_{\rm eff}} + \cdots, \\
V_{\rm b}(\phi, h) &=& \Lbr^4(h) \cos \left(\frac{\phi}{f} \right),
\eea
where $f_{\rm eff}$ is a mass scale describing the relaxion excursion\footnote{Note that we are using a different notation from  \cite{Graham:2015cka}, which appears to be more convenient for describing the physics of axion-like $\phi$. The relaxion coupling $g$ introduced in \cite{Graham:2015cka} corresponds to $g=M^2/f_{\rm eff}$ in our notation.}
  necessary to scan the Higgs mass-square $\mu_h^2$  from ${\cal O}(M^2)$ to  the final value of ${\cal O}(v^2)$,
  and  $c_0$ is a positive dimensionless coefficient which is bounded 
  as 
  \bea
  c_0 \,\gtrsim\, {\cal O}\left(\frac{1}{16\pi^2}\right)
  \eea
  to avoid a fine tuning problem. 
  With the above potential, initially the relaxion starts from a value giving  $\mu_h^2={\cal O}(M^2)>0$,  and subsequently moves to decrease $\mu_h^2(\phi)$ as enforced by the potential $V_0$.  
Eventually $\phi$  stops its motion at the value giving $\mu_h^2\simeq -(90\, {\rm GeV})^2$ due to the periodic barrier potential $V_{\rm b}$ which is developed when $\mu_h^2(\phi)$ becomes negative and  therefore $h$ gets a nonzero expectation value.

There are two different schemes to generate the barrier potential $V_{\rm b}$. 
The minimal scheme would be to generate $V_{\rm b}$ by low energy QCD  through the coupling:
\bea
 \frac{1}{32\pi^2}\frac{\phi}{f} G\tilde G,\eea where $G$ and $\tilde G$
denote the gluon field strength and its dual, respectively.
In this case, $V_{\rm b}$ corresponds to the well-known  QCD axion potential \cite{Kim:2008hd}, approximately given by 
\bea
V_{\rm b}(h,\phi) \,=\,\Lbr^4(h) \cos\left(\frac{\phi}{f}\right)\,\sim\, y_u \langle h\rangle \langle \bar u_L u_R\rangle \cos\left(\frac{\phi}{f}\right)\,\sim \,f_\pi^2 m_\pi^2\cos\left(\frac{\phi}{f}\right) ,
\eea 
where  $f_\pi$ and $m_\pi$ are the pion decay constant and the pion mass, respectively, and $y_u \sim 10^{-5}$ is the up-quark Yukawa coupling to the Higgs boson. Note that the QCD-induced $V_{\rm b}$ is linear in $h$ as it involves the $SU(2)\times U(1)$ breaking  condensation $\langle \bar u_L u_R\rangle$.

An alternative scheme is that 
$V_{\rm b}$ is generated by new physics at scales around the weak scale \cite{Graham:2015cka,Gupta:2015uea, Antipin:2015jia}.
In such case, the underlying new physics preserves the electroweak gauge symmetry, and 
the resulting barrier potential generically takes the form:
\dis{
V_{\rm b}(h,\phi) \,\equiv \,\Lbr^4(h) \cos\left(\frac{\phi}{f}\right)\,=\,
\left(\mu_0^4+ \mu_b^2|h|^2\right)\cos\left(\frac{\phi}{f}\right),
}
where $\mu_0$ and $\mu_b$ are determined by the model-dependent scale where $V_{\rm b}$ is generated, as well as the involved coupling constants \cite{Graham:2015cka,Gupta:2015uea, Antipin:2015jia}.  To implement the relaxion mechanism, one needs the Higgs-dependent part of $V_{\rm b}$ dominate over the Higgs-independent part when the relaxion is stabilized, i.e.
\dis{\mu_b^2v^2 \,> \,\mu_0^4.}
 To achieve this without a fine tuning problem,
the scale where $V_{\rm b}$ is generated should not exceed ${\cal O}(4\pi v)$, implying that the height of $V_{\rm b}$ is bounded as \cite{Graham:2015cka, Espinosa:2015eda, Antipin:2015jia, Gupta:2015uea}
\dis{\label{br_max}
\Lbr^4(h=v)  \,\sim\, \mu_b^2 v^2\,\lesssim\, {\cal O}(16\pi^2 v^4).}

Using the stationary condition  $\partial V(\phi, h)/\partial \phi = 0 $,
one can relate the relaxion excursion scale   $f_{\rm eff}$ with the other model parameters as
\dis{ \label{rel_min}
 c_0\frac{M^4}{f_{\rm eff}} \,\sim\, \frac{\Lbr^4}{f}\, \sin\left(\frac{\phi_0}{f}\right),
}
where $\phi_0$ is the vacuum value of the stabilized relaxion. As we will argue in Appendix (\ref{high_barrier}), the relaxion field is stabilized at a value yielding 
\dis{ \label{sin}
\sin\left(\frac{\phi_0}{f}\right) \sim \frac{v^2}{\Lbr^2 + v^2}.
}   
Here and
in the following, $\Lbr$ corresponds to the value when the Higgs field develops the present vacuum value  $\langle h\rangle = v =246$ GeV, i.e. $\Lbr^4 =\Lbr^4(h=v) \sim \mu_b^2v^2$.
Then the stationary condition (\ref{rel_min}) shows that the relaxion mechanism transmutes the {\it unnatural} weak scale hierarchy
$M\gg v$ to a {\it technically natural} hierarchy between the relaxion scales:
\dis{
\frac{f_{\rm eff}}{f} \,\sim\, \frac{M^4}{\Lbr^4} \, \frac{c_0}{\sin(\phi_0/f)} \,\gtrsim\,  \frac{c_0}{4\pi} \frac{M^4}{v^4},
}
where we use  (\ref{br_max}) and (\ref{sin}) for the lower bound on $f_{\rm eff}/f$.  
If the relaxion is a pseudo Nambu-Goldstone boson, both $f_{\rm eff}$ and $f$  must be interpreted as axion scales within 
the periodicity of the field variable \cite{Gupta:2015uea}. Then the above relation calls for an explanation  for the origin of the big hierarchy between the two axion scales, i.e. 
$f_{\rm eff}/f\gg 1$ which is required for $M\gg 1$ TeV.
A possible solution to this problem has been proposed in \cite{Choi:2015fiu, Kaplan:2015fuy}, inspired by the earlier works \cite{Kim:2004rp, Choi:2014rja}
based on models with multiple axions.

To implement the relaxion solution, there should be a mechanism to  dissipate away the relaxion kinetic energy which originates from the initial potential energy of ${\cal O}(c_0 M^4)$. If the energy dissipation is done by the Hubble friction during the inflationary period\footnote{For other possibilities, see Ref. \cite{Hardy:2015laa, Hook:2016mqo, Higaki:2016cqb}.}, a long relaxion excursion  requires a large number of inflationary $e$-foldings. As will be discussed in the next section, the required number of $e$-foldings severely depends on the oscillation amplitude $\Lbr^4$ of the barrier potential.
Generically  lower value of $\Lbr$ requires more $e$-foldings. As a consequence, the 
QCD-induced barrier potential requires a huge number of $e$-foldings, e.g. 
${\cal N}_e \gtrsim 10^{24}\,(M/{\rm TeV})^4$, while $V_{\rm b}$ induced by new physics around the weak scale allows the required number of $e$-foldings reduced to a much smaller value, e.g. ${\cal N}_e \sim  \,(M/{\rm TeV})^4$.

In this paper, we   first identify the relaxion parameter space for a given value of the {\it acceptable} number of inflationary $e$-foldings, which we call the cosmological relaxion window\footnote{See Ref. \cite{Kobayashi:2016bue} for a discussion from different perspective.}. We then examine observational constraints on the cosmological relaxion window.
Since a too large ${\cal N}_e$ may cause a severe fine-tuning in the inflaton sector, we will focus on the region with ${\cal N}_e\lesssim 10^{24}$, i.e. the case that  the barrier potential $V_{\rm b}$ is generated by new physics,
rather than by low energy QCD.
We find that essentially
there are three distinctive viable regions: i) a region with $f \sim \mbox{few}-200$ TeV and $m_\phi\sim 0.2-10$ GeV,
ii) another region with $ f \sim 10^6-10^9$ GeV and $m_\phi\sim  \mbox{few}-50$ MeV, and finally iii) the biggest region with $f> 10^7$ GeV and $m_\phi\lesssim 100$ eV.
The first region is particularly interesting  as it is within the reach 
of future beam dump experiment such as the SHiP \cite{Alekhin:2015byh} or improved EDM experiment such as the storage ring EDM experiment \cite{Semertzidis:2016wtd}. 
We note that these three regions include a part which allows a relatively small number of $e$-foldings less than  $10^4$, although those parts require the Higgs mass cutoff scale to be below 10 TeV.

This paper is organized as follows.
In Sec. \ref{sec_2}, we summarize the inflationary constraints on the relaxion prameters to identify
the cosmological relaxion window.
 In Sec. \ref{sec_3}, we discuss a variety of observational constraints on the relaxion window, including those from cosmological or astrophysical considerations. Sec. \ref{sec_conc} is the conclusion.

\section{cosmological relaxion window
\label{sec_2}}
In this section, we summarize the conditions for the relaxion solution to be successfully implemented, under the assumption that the initial relaxion potential energy density of ${\cal O}(c_0 M^4)$  is dissipated away by the Hubble friction during the inflationary period.
First of all, there is an upper bound on the inflationary Hubble scale $H_I$ in order for the classical motion of relaxion to be dominant over
the de-Sitter quantum fluctuation:
\dis{
\frac{\dot{\phi}}{H_I} \sim \frac{V_0^\prime(\phi)}{H_I^2} > H_I,}
implying
\dis{ \label{Hupp}
 H_I \,<\, \left(V_0^\prime(\phi)  \right)^{1/3} \sim
\left(\frac{c_0M^4}{f_{\rm eff}}\right)^{1/3} \sim \left(\frac{\Lbr}{f} \frac{v^2}{\Lbr^2+v^2}\right)^{1/3} \Lbr,
} 
where the stabilization condition (\ref{rel_min}) is used. 
Note that here $\Lbr$ corresponds to the value when the Higgs field has the present VEV, i.e. $\Lbr=\Lbr( h =v)$.

The inflationary Hubble scale  has also a lower bound coming from the condition to provide an enough friction to stop the relaxion
motion after the barrier potential $V_{\rm b}$ is generated.
Otherwise, the relaxion keeps rolling down even after the condition
(\ref{rel_min}) is satisfied because of a non-vanishing kinetic energy.
Since it takes about a Hubble time to dissipate significantly  the kinetic energy by the Hubble friction, 
this requires that the relaxion moving distance over a Hubble time be smaller than the width of the barrier potential around the time when the relaxion kinetic energy becomes comparable to the height of the barrier potential:
\dis{ \label{Hbnd_1}
\frac{\dot{\phi}}{H_I} \,\sim\, \frac{\Lbr^2}{H_I}\, \frac{v^2}{\Lambda_b^2 + v^2} \, <\, f  \frac{v^2}{\Lambda_b^2 + v^2} \quad \rightarrow\quad  H_I \,>\, \frac{\Lbr^2}{f} \,\sim\,  m_\phi.
}  

Here the factor $v^2/(\Lambda_b^2 + v^2)$ accounts for the shrinking of the barrier potential when $\Lambda_b > v$, which is explained in Appendix (\ref{high_barrier})\footnote{In fact, the barrier potential takes the form of a potential well
when $\Lbr > v$ as will be noticed in Appendix (\ref{high_barrier}). }. This bound is normally stronger than the following requirement that the inflaton energy density should be dominant over the relaxion energy density $\rho_\phi\sim c_0 M^4$:
\dis{ \label{Hbnd_2}
H_I\, >\, \sqrt{c_0}\frac{M^2}{\Mpl}.}
From (\ref{Hupp}) and (\ref{Hbnd_1}), we obtain an upper bound on the relaxion mass:
\dis{ \label{mp_max}
m_\phi \lesssim v.
}
On the other hand, (\ref{Hupp}) and (\ref{Hbnd_2}) impose an upper bound on the Higgs mass cutoff $M$ as specified later.

An important quantity for relaxion cosmology is the total number of $e$-foldings required for the relaxion to move over a field distance $\sim f_{\rm eff}$ to scan the Higgs mass from ${\cal O}(M)$ to the weak scale. For the case that the barrier potential $V_{\rm b}$ is generated by new physics,
this is estimated as 
\bea
 \label{N_NP}
 {\cal N}_{\rm NP}&\sim& \frac{f_{\rm eff}}{\dot{\phi}/H_I} 
\,\sim\,  \frac{ f_{\rm eff}^2 H_I^2}{c_0M^4} \,\sim\, \frac{c_0f^2 M^4 H_I^2}{\Lbr^8\sin^2(\phi_0/f)} 
\nonumber \\
&\gtrsim&  \max\left[\frac{1}{16\pi^2}\frac{M^4}{\Lbr^4}, \,\frac{1}{(16\pi^2)^2}\frac{f^2}{\Mpl^2}\frac{M^8}{\Lbr^8} \right]\left(1+\frac{\Lbr^2}{v^2}\right)^2,
\eea
where the stabilization conditions (\ref{rel_min}) and (\ref{sin}) are used together with  the lower bounds (\ref{Hbnd_1}) and (\ref{Hbnd_2}) on the Hubble scale, and $c_0\gtrsim 1/16\pi^2$.
Here we see that the required number of $e$-folding is minimized by $M^4/16\pi^2v^4\sim (M/{\rm TeV})^4$ for a barrier amplitude $\Lambda_b \gtrsim {\cal O}(v)$. Therefore, one can raise the Higgs mass cutoff $M$ up to
for instance 10 TeV with an inflationary $e$-folding ${\cal N}_e =  {\cal O}(10^4)$ if the barrier amplitude  is similar to or above the weak scale.


On the other hand, if $V_{\rm b}$ is generated by low energy QCD dynamics, one needs much more $e$-foldings. In fact,  in this case the scheme should be modified to avoid the strong CP problem \cite{Graham:2015cka}. 
Taking into account the inflaton-induced relaxion coupling
during inflation, which was introduced in \cite{Graham:2015cka} to avoid the strong CP problem, 
the resulting number of $e$-foldings is estimated as 
\begin{eqnarray}
{\cal N}_{\rm QCD} &\sim& \frac{1}{\theta_{\rm QCD}} \frac{c_0f^2 M^4 H_I^2}{\Lbr^8}
\,\,\gtrsim\,\, \frac{1}{\theta_{\rm QCD}}\times 
\max\left[\frac{1}{16\pi^2}\frac{M^4}{f_\pi^2 m_\pi^2},\,
\frac{1}{(16\pi^2)^2}\frac{f^2}{\Mpl^2}\frac{M^8}{f_\pi^4m_\pi^4}\right]
\nonumber \\
&\gtrsim&
\max\left[ 10^{24}\left(\frac{M}{{\rm TeV}}\right)^4,\,
10^{19}\left(\frac{f}{10^{9}\, {\rm GeV}}\right)^2\left(\frac{M}{{\rm TeV}}\right)^8\right],
\end{eqnarray}  
where we use again  the lower bounds (\ref{Hbnd_1}) and (\ref{Hbnd_2}) on the Hubble scale  with $\Lbr^2\sim f_\pi m_\pi$, together with $c_0\gtrsim 1/16\pi^2$ and
$|\theta_{\rm QCD}|\lesssim 10^{-10}$.
Although not being a rigorous argument, it is likely that
the above huge $e$-folding number  causes  a severe  fine-tuning problem in the inflaton sector 
\cite{German:2001tz, Dine:2011ws, Iso:2015wsf}.
To avoid this potential problem, in the following we will focus on the scenario that the barrier potential is generated by new physics, which allows the $e$-folding number to be much smaller than the case of QCD-induced barrier.



Requiring ${\cal N}_{\rm NP} < {\cal N}_e$ for a certain value of the acceptable $e$-folding number ${\cal N}_e$, the bound (\ref{N_NP}) is translated to
\dis{ \label{rw_M}
 M < \min\left[9\, {\rm TeV} \left(\frac{{\cal N}_e}{10^4}\right)^{1/4},\,\, 10^{11} \,{\rm GeV} 
\left(\frac{{\rm TeV}}{f}\right)^{1/6} \right] ,
}
\dis{ \label{rw_L}
 30 \, {\rm GeV} \left(\frac{M}{1\, \textrm{TeV}}\right) \left(\frac{10^4}{{\cal N}_e}\right)^{1/4}  < \Lbr \,\lesssim \, {\cal O}(\sqrt{4\pi}v),
}
\dis{ \label{rw_f}
M <  f   <  3\times10^{22} \, {\rm GeV}  \left(\frac{1\, \textrm{TeV}}{M}\right)^4  \left(\frac{\Lbr}{1\, \textrm{TeV}}\right)^4 \left(\frac{{\cal N}_e}{10^4}\right)^{1/2}  \left(1+\frac{\Lbr^2}{v^2}\right)^{-1},
}
where we assume $f>M$ for theoretical consistency,  and $\Lbr \lesssim {\cal O}(\sqrt{4\pi}v)$ to avoid a fine-tuning problem
in the new physics sector to generate the barrier potential. The second bound in (\ref{rw_M}) is derived from (\ref{Hupp}) and (\ref{Hbnd_2}).
The above parameter range corresponds to the cosmological relaxion window for the Higgs mass cutoff $M$, the barrier amplitude $\Lbr$, and the relaxion decay constant $f$, expressed in terms of the acceptable number of $e$-folding ${\cal N}_e$. 
Notice that the Higgs mass cutoff is bounded above by ${\cal O}(10)$ TeV if one requires a relatively small number of $e$-foldings smaller than ${\cal O}(10^4)$.

\begin{figure}[t]
\begin{center}
\begin{tabular}{l}
  \includegraphics[width=0.5 \textwidth]{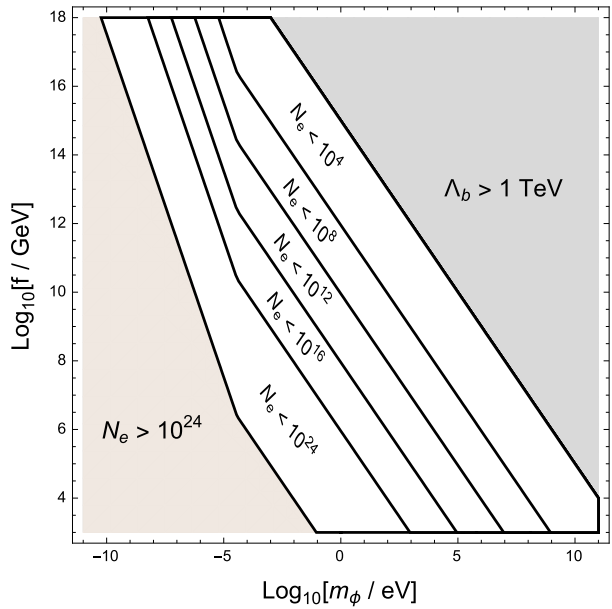}
   \end{tabular}
  \end{center}
  \caption{
Cosmological relaxion window in terms of the relaxion mass $m_\phi$ and the relaxion decay constant $f$, classified also in terms of the acceptable  $e$-folding number ${\cal N}_e$. The gray region with $\Lbr>1$ TeV  is theoretically disfavoured as it requires a fine-tuning in the underlying dynamics to generate the barrier potential.  
  }
\label{rw_overall_bare}
\end{figure}

Since the relaxion gets its mass dominantly by the barrier potential as $m_\phi \sim \Lbr^2/f$, the above relaxion window leads to
\dis{ \label{rw_mph}
 3 \times 10^{-8} \, {\rm eV} \left(\frac{M}{\textrm{TeV}}\right)^{4} \left(\frac{1\,{\rm TeV}}{\Lbr}\right)^{2} \left(\frac{10^4}{{\cal N}_e}\right)^{1/2}  \left(1+\frac{\Lbr^2}{v^2}\right) < m_\phi  < \min\left[v, \,
 1\,{\rm TeV} \left(\frac{1\,\textrm{TeV}}{M}\right)  \left(\frac{\Lbr}{1\, \textrm{TeV}}\right)^2\right],
}
where we apply the bound (\ref{mp_max}) also.
In Fig. (\ref{rw_overall_bare}), we depict the cosmological relaxion window in terms of the relaxion mass $m_\phi$ and the relaxion decay constant $f$ for the acceptable  number of $e$-folding  ${\cal N}_e\lesssim 10^{24}$ and the Higgs mass cutoff
$M>1$ TeV. The gray region with $\Lbr > 1$ TeV is theoretically  disfavoured as it requires a fine-tuning in the new physics sector to generate the barrier potential.   
In the next section, we will discuss a variety of observational constraints on this parameter region for  ${\cal N}_e < 10^{24}$, including those from cosmological and astrophysical considerations.

\section{Observational constraints 
\label{sec_3}}
In this section, we investigate phenomenological constraints on the cosmological relaxion window summarised in
(\ref{rw_M})-(\ref{rw_mph}). As argued in the previous section, one needs a new physics to generate the barrier potential
in order  for   ${\cal N}_e < 10^{24}$, which generically results in \cite{Graham:2015cka,Gupta:2015uea, Antipin:2015jia}
\dis{ \label{rel_higgs}
V_{\rm b}(\phi, h) = (\mu_0^4+ \mu_b^2 |h|^2) \cos \left(\frac{\phi}{f}\right) 
}
with $\mu_0^4 < \mu_b^2v^2$ and  $\mu_b \lesssim {\cal O}(4\pi v)$ for $v=\langle h\rangle =246$ GeV.  The new physics generating the above barrier potential  induces also the following  relaxion-photon coupling 
\dis{ \label{rel_photon}
c_{\phi \gamma} \frac{\alpha}{4\pi f} \phi F_{\mu \nu} \tilde{F}^{\mu \nu},
}
where $c_{\phi \gamma}$ is generically of order unity.\footnote{Although there exist a specific type of models yielding $|c_{\phi \gamma}|\ll 1$, e.g. the model of \cite{Gupta:2015uea}, such models should be regarded as a special case among the many possibilities which generically give $c_{\phi\gamma}={\cal O}(1)$. For instance, for the model of \cite{Gupta:2015uea}, one can consider different assignments of the global charges, which are equally well motivated as they lead to the same barrier potential, but give $c_{\phi\gamma}={\cal O}(1)$.  Another notable point is that the new physics sector generating the barrier potential typically involves some mass parameters which may need an explanation for their origin.
An attractive possibility is that those mass parameters are connected to the relaxion decay constant $f$ as in \cite{Choi:2015fiu}, for which $c_{\phi\gamma}={\cal O}(1)$ in most cases. }
As  $\mu_b$ and $f$ are constrained by the acceptable number of $e$-foldings ${\cal N}_e$, 
one can examine the phenomenological consequences of those couplings for a given range of  ${\cal N}_e$.
At any rate,
the barrier potential (\ref{rel_higgs}) provides the relaxion mass and also a relaxion-Higgs mixing,  which are estimated as
\dis{
m_\phi \sim \frac{\mu_b v}{f},
 }
\dis{ \label{mixing}
\theta_{\phi h} \sim \frac{\mu_b^2 v}{f(m_h^2 - m_\phi^2)} \, \sin\left(\frac{\phi_0}{f}\right) \sim \frac{m_\phi^2}{m_h^2 - m_\phi^2} \frac{f}{v} \left(1+\frac{f m_\phi}{v^2}\right)^{-1}.
}

Starting from (\ref{rel_higgs}) and (\ref{rel_photon}),
one can derive the effective couplings relevant for low energy  relaxion phenomenology, which  include \cite{Carmi:2012in}
\dis{\label{1pi}
& s_\theta \sum_{f} \frac{m_f}{v}  \phi\, \bar{\psi}_f \psi_f + s_\theta \frac{2m_W^2}{v} \phi W^{\mu +} W_\mu^- + s_\theta \frac{m_Z^2}{v} \phi Z^\mu Z_\mu \\
&+  s_\theta\,c_{h g} \frac{\alpha_s}{12\pi v} \phi G^{a \mu \nu} G^a_{\mu \nu} + s_\theta \, c_{h \gamma} \frac{\alpha}{4\pi v} \phi F^{\mu \nu} F_{\mu \nu} + c_{\phi \gamma} \frac{\alpha}{4\pi f} \phi F_{\mu \nu} \tilde{F}^{\mu \nu},
}
where $s_\theta=\sin \theta_{\phi h}$, $\psi_f$ denote the SM fermions, and 
\bea
c_{h g} &=& \sum_f A_f(\tau_f), \nonumber \\
c_{h \gamma} &=&  \sum_{f, \rm colors} \frac{2}{3} Q_f^2  A_f(\tau_f) -\frac{7}{2}A_v(\tau_W), \nonumber
\eea
where $\tau_i = m_\phi^2/4m_i^2$ and
\bea
A_v(\tau) &=& \frac{1}{7\tau^2}[3(2\tau-1)f(\tau) + 3\tau +2\tau^2], \nonumber \\
A_f(\tau)&=&\frac{3}{2\tau^2}[(\tau-1)f(\tau)+\tau], \nonumber \\
f(\tau) 
&=& 
\left\{ \begin{array}{ll}
(\arcsin \sqrt{\tau})^2, & \tau<1 \\
-\frac{1}{4}\left[ \ln \left( \frac{1+ \sqrt{1-\tau^{-1}} }{1- \sqrt{1-\tau^{-1}} } \right) - i \pi \right]^2. & \tau>1.
\end{array}
\right. \nonumber
\eea
Note that here we are considering a relatively simple situation  \cite{Graham:2015cka, Gupta:2015uea, Antipin:2015jia} that the relaxion does not couple to the gluon anomaly operator
$G\tilde G$, but couples to the electroweak gauge boson anomalies through the new physics sector to generate $V_{\rm b}$, and also
to the gluon kinetic operator $GG$ through the mixing with the Higgs boson. 

As we will see, in most cases of our study, the relevant relaxion mass  is in sub-GeV region. We then need the low energy relaxion couplings at scales below the QCD scale. Using the low energy realizations of the QCD operators 
 that appear in (\ref{1pi}) \cite{Leutwyler:1989xj,Shifman:1978zn,Voloshin:1986hp},  
we find the following low energy relaxion couplings to the pions, nucleons, photons and light leptons:  
\bea
\label{rel_eff}
{\cal L}_{\rm eff} &=&  2 s_\theta  \kappa  \frac{\phi}{v} \left(\frac{1}{2} \partial_\mu \pi^0 \partial^\mu \pi^0 +
\partial_\mu \pi^+ \partial^\mu \pi^- \right) - \frac{5s_\theta}{3}  \frac{\phi}{v} m_\pi^2\left(\frac{1}{2} \pi^0 \pi^0 + \pi^+ \pi^+\right) \nonumber  \\
&-& \frac{s_\theta}{6}   \frac{g_2 m_N}{m_W} \phi \bar{N} N + s_\theta   \frac{c_{h \gamma}\alpha}{4\pi v} \phi F^{\mu \nu} F_{\mu \nu} +  \frac{c_{\phi \gamma}\alpha}{4\pi f} \phi F_{\mu \nu} \tilde{F}^{\mu \nu} + s_\theta \sum_{l = e, \, \mu} \frac{m_l}{v}  \phi\, \bar{\psi}_l \psi_l ,
\eea
where
\bea
c_{h \gamma} \,=\,   \sum_{f, \rm colors} \frac{2}{3} Q_f^2  A_f(\tau_f) -\frac{7}{2}A_v(\tau_W) +\frac{8}{27}\,=\, 0.1-1.4 \quad
\mbox{for} \,\, m_\phi<1\,\,{\rm GeV}.
\eea

\begin{figure}[t]
\begin{center}
 \begin{tabular}{l}
  \includegraphics[width=0.45 \textwidth]{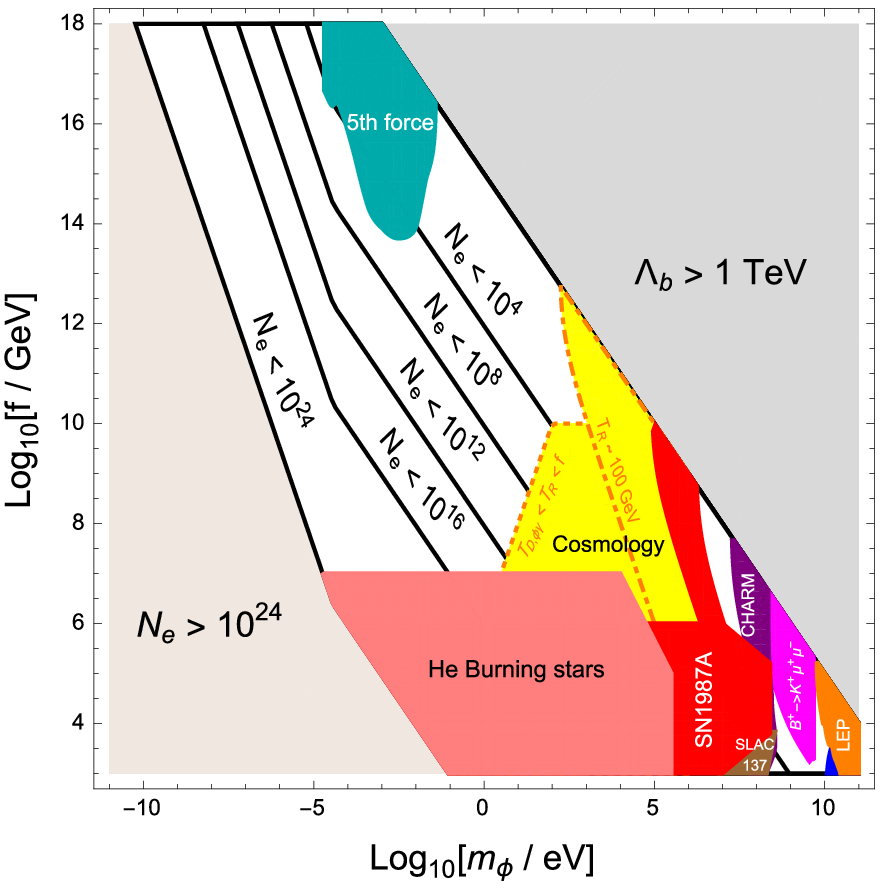}
  \hspace{0.3cm}
  \includegraphics[width=0.45 \textwidth]{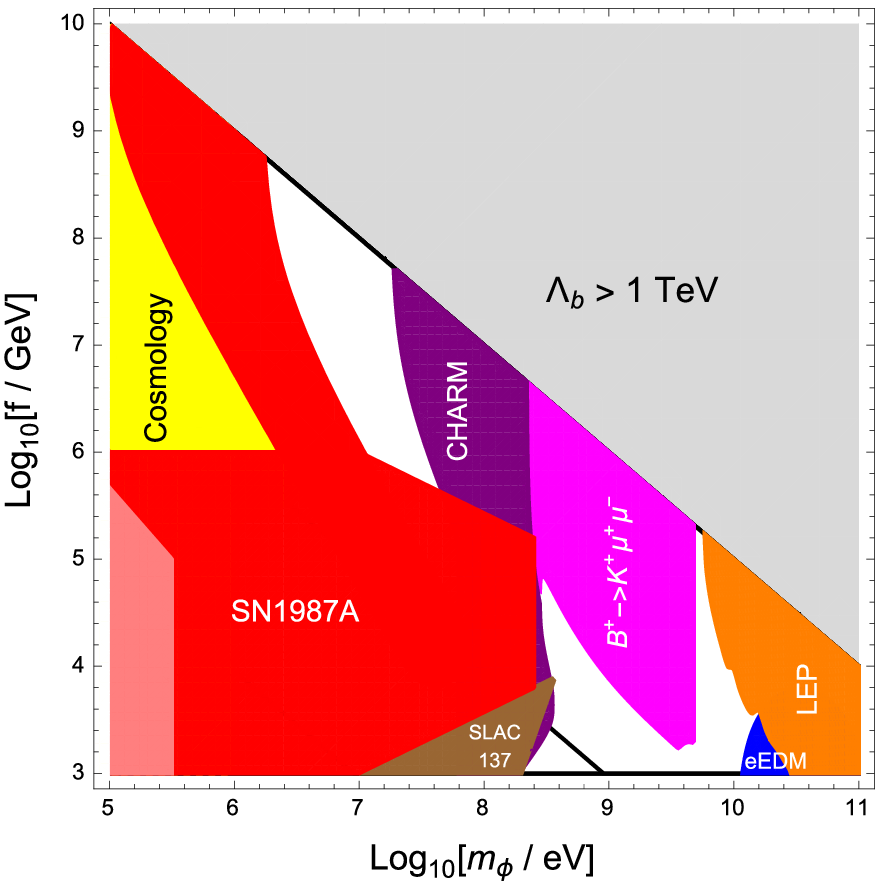}
   \end{tabular}
  \end{center}
  \caption{
Cosmological relaxion window with colored regions excluded by the observational constraints discussed in this paper. The yellow region from cosmology depends on the reheating temperature $T_R$ and shrinks for smaller $T_R$. Here we set $c_{\phi\gamma}=1$ and depict the results for $T_R\sim f$ and 100 GeV.}
\label{rw_overall}
\end{figure}

We then apply the above relaxion effective interactions to various low energy processes as
described below. The result is summarized in Fig. (\ref{rw_overall}).  Colored region
in the figure  is excluded by the constraints discussed here. 
The yellow region from cosmological considerations depends on the reheating temperature, and shrinks for lower
 reheating temperature.
One can see that essentially
there are three distinct viable regions:
i) a window with $f \sim \mbox{few}-200$ TeV and $m_\phi\sim 0.2-10$ GeV,
ii) another window with $ f \sim 10^6-10^9$ GeV and $m_\phi\sim \mbox{few}-50$ MeV, and finally iii) the biggest window with $f> 10^7$ GeV and $m_\phi\lesssim 100$ eV.

\begin{figure}[h]
\begin{center}
 \begin{tabular}{l}
 \includegraphics[width=0.5 \textwidth]{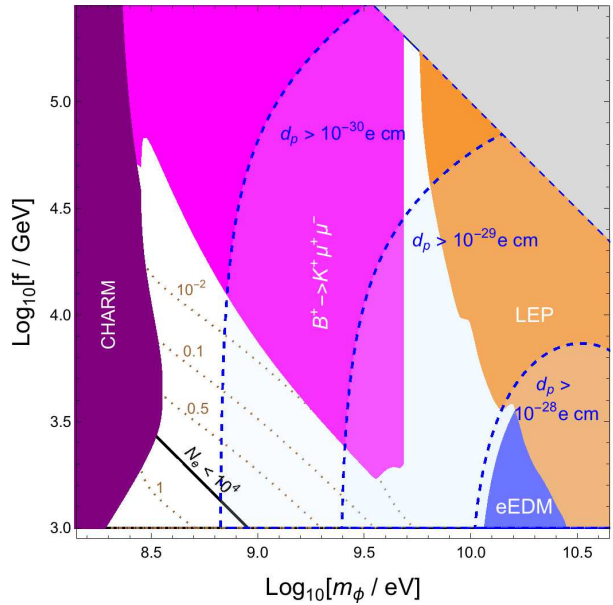}
   \end{tabular}
  \end{center}
  \caption{
Enlarged picture for the first viable window with $f \lesssim 200$ TeV.
The dashed blue lines represent the proton EDM $d_p = 10^{-28}, \,10^{-29}, \,10^{-30}\,e \cdot \rm cm$ for $c_{\phi\gamma}=1,$ respectively. 
The dotted brown line denotes the branching fraction of the relaxion decay into 2 photons. 
The allowed region is in reach of the projected SHiP experiment  ($m_\phi \lesssim$ 5 GeV) and future electron/proton EDM sensitivity.
  }
\label{rw_small}
\end{figure}

Among these three regions, the first window is particularly interesting as it is within the reach of near future experiments.
Enlarged picture of this region is depicted in Fig. (\ref{rw_small}). 
For the parameter space of $m_\phi \lesssim 3$ GeV in this region, relaxions decay dominantly into photons, and also into
 muons or pions with comparable branching ratio, which allows the parameter space
 probed by the SHiP experiment \cite{Alekhin:2015byh}. This region can be probed also by the
 future storage ring EDM experiment \cite{Semertzidis:2016wtd} which is claimed to improve the present bounds on the nucleon EDMs   by several orders of magnitudes.
In the following, we provide a description for the details of the constraints depicted in Fig. (\ref{rw_overall}). 

\subsection{LEP}
The relaxion with a mass between 5 GeV and 100 GeV is mostly constrained by the LEP experiment
through the process  $e^+ e^- \rightarrow Z \rightarrow Z^* \phi$ or $e^+ e^- \rightarrow Z^* \rightarrow Z \phi$
as noticed in \cite{Flacke:2016szy}, where the $ZZ \phi$ coupling arises from the relaxion-Higgs mixing, and $\phi$ subsequently decays to the SM particles with the same branching ratios as the corresponding SM Higgs boson of an equal mass.\footnote{The relaxion-photon coupling $c_{\phi \gamma}$ in (\ref{rel_photon}) can change the branching ratios when the mixing angle $\theta_{\phi \gamma}$ is very small. Still, it turns out that the mixing angle is large enough to suppress the photon branching ratio over the relevant mass region.} The LEP experiment provides an upper bound on the cross section of the processes normalized to the value of the SM Higgs boson depending on the Higgs-like particle's mass (here, relaxion). This is translated to an upper bound on $\sin^2 \theta_{\phi h}$ in terms of $m_\phi$. As one can see from the relaxion-Higgs mixing (\ref{mixing}), the upper bound on the mixing angle gives an upper bound on $f$ for a given $m_\phi$.
 
The former process with an on-shell intermediate $Z$ boson, which is analyzed by L3 \cite{Acciarri:1996um}, imposes the most stringent bound on the mixing angle for a relaxion mass below about 30 GeV. For a larger mass up to 116 GeV, the four LEP collaborations ALEPH, DELPHI, L3, and OPAL provide a bound on the cross section of the latter process with a final on-shell $Z$ boson \cite{Schael:2006cr}. 

In Fig. (\ref{rw_small}), we see that the LEP constraints exclude a relaxion heavier than 30 GeV within the relaxion window, while constraining the relaxion decay constant for a relaxion mass between 5 GeV and 30 GeV.
We remark that the LHC bound concerning the Higgs decay to two relaxions $h\rightarrow \phi \phi$ strongly constrains the relaxion-Higgs mixing angle beyond the LEP for $m_\phi \gtrsim 25$ GeV \cite{Flacke:2016szy}. However, this mass region is almost excluded already  by the LEP and electron EDM bounds within the relaxion window as one can find in Fig. (\ref{rw_small}). 

\subsection{EDM}
\begin{figure}[t]
\begin{center}
 \begin{tabular}{l}
  \includegraphics[width=0.5 \textwidth]{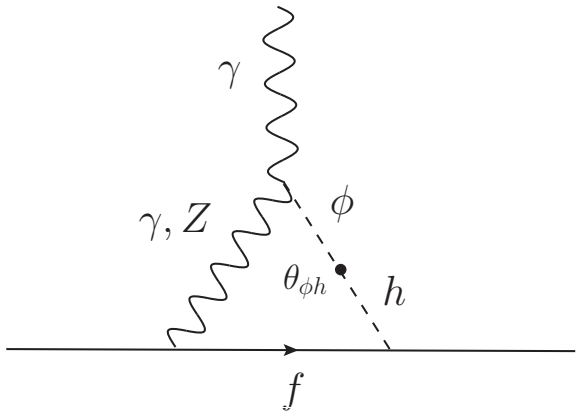}
   \end{tabular}
  \end{center}
  \caption{
EDM of light fermion from the relaxion mixing with the SM Higgs boson.
  }
\label{eEDM}
\end{figure}

A simultaneous presence of the relaxion-Higgs mixing and the relaxion-photon coupling  $\phi F\tilde F$ violates 
the CP invariance, so can induce nonzero electric dipole moments (EDMs). For instance, EDMs of  light fermions arise from the diagram of Fig. (\ref{eEDM}), yielding \cite{Jung:2013hka, Marciano:2016yhf}
\dis{
d_f \,\simeq\, 4 \frac{e^3}{(4\pi)^4} \frac{m_f}{v} \frac{c_{\rm \phi \gamma}}{f} \sin\theta_{\phi h} \cos\theta_{\phi h} \ln \left(\frac{m_h^2}{m_\phi^2}\right).}
Applying this to the electron EDM, we find
\dis{d_e\,\sim\, 7\times 10^{-29} c_{\phi\gamma} \left(\frac{m_\phi}{10\,{\rm GeV}}\right)^2 \ln\left(\frac{10 \, {\rm GeV}}{m_\phi}\right) \left(1+\frac{f m_\phi}{v^2}\right)^{-1}\, {e \cdot \rm cm}.
}
The current experimental bound on the electron EDM is $d_e \, < \, 8.7 \times 10^{-29}\, e \cdot \rm cm$  \cite{Baron:2013eja}. 
This implies that  $m_\phi \gtrsim  10/\sqrt{c_{\phi\gamma}}$ GeV is excluded if the relaxion decay constant $f$ is below $v^2/m_\phi \sim 10 \sqrt{c_{\phi\gamma}}$ TeV. This constraint from the electron EDM is depicted in Fig. (\ref{rw_overall}) under the assumption that $c_{\phi\gamma}=1$.  Our result suggests that the relaxion with a mass below 10 GeV can be probed further by future EDM experiments, particularly by the storage ring EDM experiment which is claimed to improve the bound on the proton EDM down to $d_p \sim 10^{-29}\, e \cdot \rm cm$  \cite{Semertzidis:2016wtd} with a final goal $d_p \sim 10^{-30}\, e \cdot \rm cm$ \cite{Semertzidis}. In the enlarged Fig. (\ref{rw_small}), we depict also the parameter region yielding the proton EDM $d_p = 10^{-28}, \,10^{-29}, \,10^{-30}\,e \cdot \rm cm$ for $c_{\phi\gamma}=1.$
Here the proton EDM is estimated by applying the QCD sum rule with the following relation \cite{Hisano:2015rna}:
\dis{
d_p = 0.78 \,d_u (\mu_*) -0.20 \, d_d (\mu_*),
}
where the renormalization scale is taken to be $\mu_* = 1$ GeV.\footnote{
If we use the the Naive Dimensional Analysis \cite{NDA} assuming that strange quark contribution is dominant, the resultant proton EDM turns out to 
be larger by an order of magnitude.}

\subsection{Rare meson decay}
For the relaxion with a mass below 5 GeV having a nonzero relaxion-Higgs mixing, there are strong constraints coming  from rare meson decays \cite{Clarke:2013aya}. The strongest one turns out to be $B^+ \rightarrow K^+ \phi \,(\phi \rightarrow \mu^+ \mu^-)$ for $m_\phi > 2m_\mu$.
Using the results from the $B$ factories \cite{Aubert:2008ps, Wei:2009zv} and the SM prediction ${\rm Br}(B^+ \rightarrow K^+ \mu^+ \mu^-)_{\rm SM} \gtrsim 2.3 \times 10^{-7}$, the new physics contribution  is constrained as \cite{Clarke:2013aya}
\dis{ \label{BtoK}
{\rm Br}(B^+\rightarrow K^+ \phi)  \times {\rm Br}(\phi \rightarrow \mu^+ \mu^-) 
\times \int_{0}^{\pi} d\theta \frac{\sin \theta}{2}\left(1-\exp\left[\frac{-l_{\rm min}}{\sin \theta}\frac{\Gamma_\phi}{\gamma}\right]\right) \lesssim 3\times 10^{-7}
}
where $l_{\rm min} \simeq 25$ cm  is the transverse decay distance from the beampipe
\cite{Batell:2009jf} and $\gamma \simeq m_B/(2m_\phi)$. The branching ratio
for $B\rightarrow K\phi$  is calculated to be  \cite{Batell:2009jf}
\dis{
{\rm Br}(B^+ \rightarrow K^+ \phi) \approx \sin^2\theta_{\phi h} \times 0.5 \times \frac{\sqrt{m_B^4 -2(m_K^2+m_\phi^2) m_B^2 + (m_K^2 - m_\phi^2)^2}}{m_B^2} \times {\cal F}_K^2(m_\phi) \nonumber
}
with the form factor ${\cal F}_K^2(m_\phi) = (1-m_\phi^2/38\,{\rm GeV}^2)^{-1}$ \cite{Ball:2004ye}.
On the other hand, the total decay width of relaxion is given by
\dis{
\Gamma_\phi = \Gamma(\phi \rightarrow \gamma\gamma) +  \Gamma(\phi \rightarrow e^+ e^-) + \Gamma(\phi \rightarrow \mu^+ \mu^-) + \Gamma(\phi \rightarrow \pi \pi), 
}
where
\bea
\Gamma(\phi \rightarrow \gamma\gamma) &=& \frac{1}{4\pi}\left(\frac{\alpha}{4\pi}\right)^2 \frac{m_\phi^3}{f^2} \left(|c_{\phi \gamma}|^2+|c_{h\gamma}|^2\frac{f^2}{v^2} \sin^2 \theta_{\phi h} \right), \label{gg} \\
\Gamma(\phi \rightarrow l^+ l^-) &=& \sin^2 \theta_{\phi h} \times \frac{m_\phi}{8\pi}  \frac{m_l^2}{v^2} \left(1-\frac{4m_l^2}{m_\phi}\right)^{3/2}, \\
 \Gamma(\phi \rightarrow \pi \pi) &=& R_{\pi \mu} \times \Gamma(\phi \rightarrow \mu^+ \mu^-), \label{pi_mu}
\eea
with  $R_{\pi \mu}$ which can be calculated\footnote{In fact, there can be a sizable uncertainty in the  value of $R_{\pi \mu}$ as discussed in \cite{Voloshin:1985tc, Raby:1988qf, Truong:1989my, Donoghue:1990xh}. This can lead to a factor 2-3 difference in the resultant upper bound on the relaxion decay constant $f$. } \cite{Voloshin:1985tc} using the effective interactions (\ref{rel_eff}):
\dis{
R_{\pi \mu} = \frac{1}{27} \frac{m_{\phi}^2}{m_\mu^2} \left(1+\frac{11}{2} \frac{m_\pi^2}{m_\phi^2}\right)^2 \frac{\beta_\pi}{\beta_\mu^3} \qquad\left(
\beta_x \equiv (1-4m_x^2/m_\phi^2)^{1/2}\right).
}
At the end, the constraint (\ref{BtoK}) turns out to put an upper limit on the relaxion-Higgs mixing angle:
\dis{
{\rm Br}(\phi \rightarrow \mu^+ \mu^-) \times \sin^2 \theta_{\phi h} \lesssim 6\times10^{-7},
} 
leading to 
\dis{
f \,\lesssim\,  7\, {\rm TeV} \left(\frac{1 \, {\rm GeV}}{m_\phi}\right)^2 \left(\frac{0.2}{\textrm{Br}_{\phi\rightarrow \mu\mu}}\right)^{1/2}  \left(1+\frac{f m_\phi}{v^2}\right)
~~ {\rm for} ~~ 2m_\mu < m_\phi \lesssim 5 \,{\rm GeV}.
}

\subsection{Beam dump experiments}
For a relaxion mass around or below $2m_\mu$, the bound from the CHARM beam dump experiment \cite{Bergsma:1985qz} becomes important. Following \cite{Bezrukov:2009yw, Clarke:2013aya}, the number of decaying relaxions inside the detector can be estimated as
\dis{
{N_{\phi, \rm d}} \,\simeq\, {N_{\phi, 0}} \left[ {\rm Br}(\phi \rightarrow \gamma \gamma,  e^+ e^-,\mu^+ \mu^-) \right] 
\left( e^{- \Gamma_\phi \frac{l_{\rm d}}{\gamma}} - e^{- \Gamma_\phi \frac{l_{\rm d}+\tilde l_{\rm d}}{\gamma}} \right),
}
where $\gamma \simeq 10 \,{\rm GeV}/m_\phi$, $l_{\rm d}=480\,{\rm m}$ is the detector distance from the target, $\tilde l_{\rm d}=35\,{\rm m}$ is the length of the detector, and the total number of produced relaxions  is estimated as
\bea
N_{\phi, 0} \,\approx \, 2.9 \times 10^{17} \frac{\sigma_\phi}{\sigma_{\pi^0}} \eea
with
\bea
 \frac{\sigma_\phi}{\sigma_{\pi^0}} &\approx& 3 \left[ \chi_s \times \frac{1}{2} {\rm Br} (K^+ \rightarrow \pi^+ \phi) +
 \chi_s \times \frac{1}{4} {\rm Br} (K_L \rightarrow \pi^0 \phi) + \chi_b \times {\rm Br} (B \rightarrow X \phi) \right],\nonumber 
\eea  
where $\chi_s = 1/7$,  $\chi_b=3\times 10^{-8}$, and the branching fractions are
given by \cite{Leutwyler:1989xj, Grinstein:1988yu}, 
\bea
{\rm Br}(K^+ \rightarrow \pi^+ \phi) &\approx& \sin^2\theta_{\phi h} \times 0.002 \times \frac{\sqrt{m_K^4 -2(m_\pi^2+m_\phi^2) m_K^2 + (m_\pi^2 - m_\phi^2)^2}}{m_K^2}, \nonumber \\
{\rm Br}(K_L \rightarrow \pi^0 \phi) &=& {\rm Br}(K^+ \rightarrow \pi^+ \phi) \times \frac{\Gamma(K^+)}{\Gamma(K_L)}, \nonumber \\
{\rm Br}(B \rightarrow X \phi) &\approx& \sin^2\theta_{\phi h} \times 0.26 \times \left(\frac{m_t}{m_W}\right)^4\left(1-\frac{m_\phi^2}{m_B^2}\right)^2. \nonumber
\eea

The result of the CHARM experiment requires that $N_{\phi, \rm d} < 2.3$ at $90\%$ C.L. Since $N_{\phi, \rm d}$ is roughly proportional to $ (m_\phi^4 f^2/v^6)\exp\left[- m_\phi^6 m_l^2/v^8 \times (f/{\rm eV})^2 \right]$, it excludes a certain range of $f$ for a given $m_\phi$.  
On the other hand, the SLAC 137 beam dump experiment \cite{Bjorken:1988as} excludes some region with $m_\phi<100$ MeV and $f <$ 10 TeV, which results from the relaxion-photon coupling (\ref{rel_photon}).

\subsection{Cosmological constraints}
Relaxion may affect the
Big Bang Nucleosythesis  or the Cosmic Microwave Background (CMB). It
may also contribute to dark matter, dark radiation, extragalactic background lights, or galatic X-Rays, depending on the relaxion mass and lifetime \cite{Cadamuro:2011fd}.
The bounds from these considerations depend on the amount of relaxions produced in the early universe, which in turn depends on the reheating temperature.

If the reheating temperature is large enough,  relaxions will be 
in thermal equilibrium by the relaxion-photon coupling (\ref{rel_photon}).   
The decoupling temperature for the coupling (\ref{rel_photon}) turns out to be \cite{Cadamuro:2011fd}
\dis{
T_{D, \phi\gamma} \simeq 100 \,{\rm GeV} \left(\frac{f}{c_{\phi\gamma}\times  10^6 {\rm \,GeV}}\right)^2.}
Therefore, for $f > c_{\phi\gamma}\times  10^6$ GeV, relaxions cannot be in thermal equilibrium by the relaxion-photon coupling alone, unless the reheating temperature is substantially larger than the weak scale. 
However relaxion couplings resulting from the mixing with the SM Higgs boson can make relaxions in thermal equilibrium  even when $f > c_{\phi\gamma}\times 10^6$ GeV and $T\lesssim 100$ GeV.
The dominant process for equilibrium is the single relaxion production through the collisions of SM particles, SM$+$SM $\rightarrow \phi + g$, where $g$ denotes the gluons. The thermal averaged cross section of this process is estimated as $\langle \sigma_A \beta \rangle\sim (m_f/v)^2\, \sin^2\theta_{\phi h}\,/\,T^2$, where $\beta$ is the relative velocity of the colliding two SM particles. Then relaxions are in thermal equilibrium if
\dis{
m_f \,\lesssim\, T \,\lesssim\, \left(\frac{m_f}{v}\right)^2\,\sin^2\theta_{\phi h}\, \Mpl \sim
\left(\frac{m_f}{v}\right)^2 \left(\frac{m_\phi}{m_h}\right)^4 \left(\frac{f}{v}\right)^2 \left(1+\frac{f m_\phi}{v^2}\right)^{-2}\Mpl,
}
which requires
\dis{ \label{cosmo_f}
m_\phi \gtrsim 10^5 \,{\rm eV} \left(\frac{v}{m_f}\right)^{1/4} \left(\frac{10^6 \,{\rm GeV}}{f}\right)^{1/2} \left(1+\frac{f m_\phi}{v^2}\right)^{1/2}. 
}
If the reheating temperature $T_R$ is greater than the electroweak scale so that the top quark interaction can be effective
in (\ref{cosmo_f}) with $m_f = m_t$, the relaxion is efficiently produced from the thermal bath  for large region of $f$ as far as the relaxion is heavy enough.

In Fig. (\ref{rw_overall}), we show the excluded parameter region for two different choices of the reheating temperature: $T_{D, \phi \gamma} < T_R < f$ (dotted) and $T_R \sim 100$ GeV (dot-dashed). 
Obviously the excluded region shrinks as the reheating temperature becomes smaller. 

\subsection{SN1987A and He Burning stars}
The SN1987A energy loss argument and the life time of helium burning stars can give a stringent bound on the relaxion decay constant $f$. 
Here we assume $c_{\phi\gamma}={\cal O}(1)$, and take the results of  \cite{Masso:1995tw}, which 
are based on the Primakoff process due to the CP conserving relaxion coupling (\ref{rel_photon}) to $F\tilde F$. The CP violating relaxion-photon coupling $\phi FF$ induced by the relaxion-Higgs mixing  is negligible over the relevant parameter space. 

The relaxion-nucleon coupling $\phi\bar N N$ in (\ref{rel_eff}), which originates from the relaxion-Higgs mixing, gives rise to an additional constraint through the relaxion emission by the nucleon-nucleon bremsstrahlung process, which has been studied in \cite{Krnjaic:2015mbs, Ishizuka:1989ts}.\footnote{See \cite{Giannotti:2005tn} for the constraints associated with the CP conserving ALP-nucleon couplings
of the form $\partial_\mu \phi \bar N\gamma^\mu \gamma_5 N$.} Applying the results of \cite{Krnjaic:2015mbs, Ishizuka:1989ts} to the relaxion case, we find that some of the region with
the relaxion decay constant in the range $10^6 \, {\rm GeV} \lesssim f \lesssim 10^{10}\, {\rm GeV}$  is further excluded for the relaxion mass in the range $0.1 \, {\rm MeV} \lesssim m_\phi \lesssim 10\, {\rm MeV}$.

\subsection{The 5th force}

A light relaxion can mediate a long range force through the Yukawa couplings to the SM particles induced by the relaxion-Higgs  mixing \cite{Flacke:2016szy}. 
Since the resulting Yukawa couplings do not exactly scale with the masses, this force violates the equivalence principle.  
At the Newtonian approximation, the total effective gravitational potential between two bodies $A$ and $B$ including the relaxion mediated force can be written as,
\dis{
V(r) = - G_N \frac{m_A m_B}{r} \left(1 + \tilde{\alpha}_A \tilde{\alpha}_B e^{-m_\phi r}\right).
}
The couplings $\tilde{\alpha}_{A,B}$ are  given by the sum of the universal contribution from the nucleons and a subleading element-dependent part. The universal part is calculated to be \cite{Piazza:2010ye}   
\dis{
\tilde{\alpha} = c_{\phi N} \frac{\sqrt{2} \Mpl}{m_N},
}
where $\Mpl$ is the reduced Planck mass, and
\dis{
c_{\phi N} =  \frac{g_2 m_N}{6\, m_W} \, \sin \theta_{\phi h}
}
which is the $\phi \bar{N} N$ coupling found in (\ref{rel_eff}). On the other hand, the subleading element-dependent part leads to a variation of acceleration depending on the test bodies.  One can then put an upper bound on the universal coupling $\tilde{\alpha}$ 
from the torsion balance experiment \cite{Schlamminger:2007ht} testing the equivalence principle, which in turn constrains the relaxion-Higgs mixing angle $\theta_{\phi h}$. The relevant interaction length ranges from $10^{-2}$ m to a very long distance over $10^{12}$ m, which corresponds to a relaxion mass below  $10^{-5}$ eV.  
However, it turns out that the coupling $\tilde{\alpha}$ within the relaxion window is fairly small compared to the  experimental upper bounds if the relaxion decay constant $f$ is sub-Planckian.

For a relaxion mass from $10^{-5}$ eV to $0.1$ eV (i.e. the interaction length from $10^{-2}$ m to $10^{-6}$ m), various experimental tests of the gravitational inverse-square law constrains the universal coupling $\tilde{\alpha}$ depending on the interaction length \cite{Kapner:2006si}. This restricts the relaxion-Higgs mixing considerably, so that it excludes some of the parameter region with  $f\gtrsim 10^{14}$ GeV, as depicted by the green colored region of Fig. (\ref{rw_overall}). 

For a larger relaxion  mass above $0.1$ eV, the bounds from the Casimir effect \cite{Bordag:2001qi} and neutron scattering experiment \cite{Nesvizhevsky:2007by} might be relevant for constraining the relaxion-mediated force \cite{Flacke:2016szy}. However, 
they turn out to be too weak to exclude any of the parameter region within the relaxion window.

\section{Conclusion} \label{sec_conc}

To implement the relaxion solution to the weak scale hierarchy problem, there should be a mechanism to dissipate away the initial relaxion potential energy of ${\cal O}(c_0 M^4)$, where $M$ is the Higgs mass cutoff scale presumed to be well above the weak scale and $c_0 \gtrsim {\cal O}(1/16\pi^2)$ to avoid a fine tuning problem.
One typically assumes that the required dissipation of relaxion energy is achieved by the Hubble friction during the inflationary period. Then  the scheme requires a rather large number of inflationary $e$-foldings which may cause a fine tuning problem in the inflaton sector.
In the minimal scenario that the barrier potential is generated by low energy QCD dynamics, the required $e$-folding number is huge, ${\cal N}_e\gtrsim 10^{24}(M/{\rm TeV})^4$.
On the other hand, in the alternative scenario that the barrier potential is generated by new physics around the weak scale, the required $e$-folding number can be greatly reduced, e.g. ${\cal N}_e\gtrsim  (M/{\rm TeV})^4$.

In this paper, we classified the parameter space of the relaxion mass $m_\phi$ and the decay constant $f$ in terms of a given value of the acceptable $e$-folding number, and examine a variety of observational constraints on the parameter region with ${\cal N}_e \lesssim 10^{24}$.
After taking into account the observational constraints discussed in this paper, three viable windows survive:
i) a window with $f \sim \mbox{few}-200$ TeV and $m_\phi\sim 0.2-10$ GeV,
ii) another window with $ f \sim 10^6-10^9$ GeV and $m_\phi\sim \mbox{few}-50$ MeV, and finally iii) the biggest window with $f> 10^7$ GeV and $m_\phi\lesssim 100$ eV.
The first window is particularly interesting  as it is within the reach 
of future beam dump experiment such as the  SHiP experiment \cite{Alekhin:2015byh}   
 or improved EDM experiment such as the storage ring EDM experiment \cite{Semertzidis:2016wtd}. 
The parameter region with $f\,>10^6$ GeV is constrained by a variety of cosmological/astrophysical bounds depending on the reheating temperature. All three windows include a parameter region with relatively small number of $e$-foldings less than $10^4$, although such region requires the Higgs mass cutoff scale to be below 10 TeV.
 
\section{Acknowledgment}
We thank Gordan Krnjaic for informing us the previous works on the CP-even scalar emission from SN1987A, and Surjeet Rajendran for informing the potential importance of the 5th force constraints.
While revising the first version, we are indebted to Ref. \cite{Flacke:2016szy} for valuable informations on the LEP/LHC bound for the high relaxion mass region $m_\phi > 5$ GeV, and also the constraints from the 5th force. We thank Thomas Flacke and Gilad Perez for the communications related to Ref. \cite{Flacke:2016szy}.
This work was supported by IBS under the project code, IBS-R018-D1 [KC and SHI], and by the German Science Foundation (DFG)
within the SFB-Transregio TR33 ``The Dark Universe" [SHI].

\appendix

\section{Relaxation with a barrier amplitude bigger than the weak scale} \label{high_barrier}

In this appendix, we discuss the relaxion stabilization procedure when the amplitude of the oscillating barrier potential is bigger than the weak scale.\footnote{See Ref. \cite{Flacke:2016szy} for an argument that the oscillation amplitude of the barrier potential is bounded above by the weak scale. Here we are pointing out an alternative possibility.}
In this case, the barrier potential takes part in scanning the Higgs mass as
\dis{
V(\phi, h) = \mu_h^2 (\phi) |h|^2 - c_0 M^4 \frac{\phi}{f_{\rm eff}} + \cdots,
}
where
\dis{
\mu_h^2 (\phi) = M^2 - M^2 \frac{\phi}{f_{\rm eff}} + \mu_b^2 \cos\left(\frac{\phi}{f}\right) + \cdots.
}

For $v< \mu_b \ll M$, $\mu_h^2$ is initially positive and of ${\cal O}(M^2)$. As the relaxion rolls down, it arrives eventually  at the point where  $M^2 - M^2 \phi/f_{\rm eff}=\mu_b^2 - \Delta^2$ for  $0< \Delta^2 < \mu_b^2$. Then $\mu_h^2$ oscillates between $2\mu_b^2 - \Delta^2$ and $- \Delta^2$ as the relaxion moves over the period $2\pi f$,  and a non-zero Higgs VEV $\langle h\rangle \sim \Delta$ is generated when $\mu_h^2\sim -\Delta^2$. Note that
 $\Delta^2$ is increasing by $M^2 f/f_{\rm eff} \sim (\Lbr^4/M^2) \sin(\phi_0/f) \ll v^2$ as the relaxion moves over a distance
   $\sim f$, and therefore it can be finely scanned. 
 In this case, a non-zero barrier potential is developed only over a narrow range of the relaxion field
near the point of $\cos(\phi_*/f) = -1$,  and therefore takes the form of quadratic potential well
 with a width $\Delta\phi \sim f\Delta/\mu_b$ and a depth $\Delta V_{\rm b} \sim \Lbr^4 \Delta^2/\mu_b^2$.
 
 As we have noticed in (\ref{Hbnd_1}), the relaxion can be successfully stabilized by this potential well when $\Delta^2 \sim v^2$,  if the inflationary Hubble scale $H_I > m_\phi$.
The width and depth of the potential well  stabilizing the relaxion is suppressed  by $v/\mu_b$ and $v^2/\mu_b^2$, respectively, so are given by
\bea
\Delta\phi &\sim& \frac{v}{\mu_b}f \,\sim\, \frac{v^2}{\Lbr^2+v^2}f,
 \\
\Delta V_{\rm b} &\sim& \frac{v^2}{\mu_b^2}\Lbr^4 \,\sim\, v^4,
\eea
and  $\sin(\phi/f)$ for the stabilized relaxion  is bounded as
\dis{ \label{sin_phi}
\sin\left(\frac{\phi_0}{f}\right) \lesssim \frac{v}{\mu_b} \sim \frac{v^2}{\Lbr^2+v^2},
}
where $\Lbr^2 = \mu_b v$ as defined in (\ref{br_max}). 
Generically it takes about a Hubble time to dissipate significantly the relaxion kinetic energy by the Hubble friction.  
For $H_I > m_\phi$,  the relaxion moving distance over a single  Hubble time is smaller than the width of the potential well, as discussed in (\ref{Hbnd_1}).
As a result, the relaxion kinetic energy from the potential well can be efficiently dissipated away by the Hubble friction, which makes the relaxion eventually stabilized within the potential well.

\end{document}